\documentclass[aps,pra,twocolumn,groupedaddress,showpacs]{revtex4-1}
\usepackage{graphicx}
\usepackage{amsfonts}

\begin{document}

\preprint{1.1}

\title{Prospects for rapid deceleration of small molecules by optical bichromatic forces}

\author{M. A. Chieda}
\author{E. E. Eyler}
\affiliation{Physics Department, University of Connecticut, Storrs, CT 06269}

\date{\today}

\begin{abstract}
We examine the prospects for utilizing the optical bichromatic force (BCF) to greatly enhance laser deceleration and cooling for near-cycling transitions in small molecules.  We discuss the expected behavior of the BCF in near-cycling transitions with internal degeneracies, then consider the specific example of decelerating a beam of calcium monofluoride molecules.  We have selected CaF as a prototype molecule both because it has an easily-accessible near-cycling transition, and because it is well-suited to studies of ultracold molecular physics and chemistry.  We also report experimental verification of one of the key requirements, the production of large bichromatic forces in a multi-level system with significant energy level splittings, by performing tests in an atomic beam of metastable helium.
\end{abstract}


\maketitle

\section{Introduction}

Despite the immense successes of laser deceleration and cooling for atoms, the direct application of these methods to molecules has been extremely limited, due to the absence of true two-level cycling transitions even in the simplest of diatomic molecules.  Instead, the explosive recent growth of activity on cold and ultracold molecules has relied on indirect methods including photoassociation, magnetoassociation, buffer-gas cooling, and others \cite{Carr09}.  The sole exception is the very recent demonstration of 1-D cooling and laser deceleration for SrF molecules in the near-cycling $A-X$ transition \cite{Shuman09,Shuman10,DeMille11}.  This was accomplished by using numerous repumping frequencies to recover molecules lost by radiation to the wrong vibrational, fine-structure, or hyperfine level, thus permitting the thousands of excitation-decay cycles required to achieve a significant radiative force.

In this paper we examine the prospects for greatly enhancing the radiative force on molecules by utilizing the optical bichromatic force (BCF), in which many stimulated cycles of excitation and emission occur for each radiative cycle.   There are numerous diatomic molecules with near-cycling transitions for which laser deceleration and cooling should be at least marginally practical \cite{DiRosa04,Stuhl08,Shuman09}, many of them polar molecules that are of particular interest at ultracold temperatures because of their strong anisotropic interactions \cite{Carr09, Krems09}.  If a BCF enhancement of the radiative force by 1-2 orders of magnitude can be obtained, the task becomes far easier.  As a representative example, we examine in some detail the requirements for decelerating a beam of calcium monofluoride molecules.  We select CaF as a prototype molecule both because it has an easily-accessible near-cycling transition and because it is well-suited to studies of ultracold molecular physics and chemistry.  Its excited-state spectrum has been studied very extensively, particularly in Rydberg states \cite{Field75,Bernath80,Berg93,Gittins01,Kay04,Field05,Kay08,Petrovic09,Kay11,Pelegrini05,Wall08}, which should greatly facilitate future high-resolution experiments on laser-cooled beams or trapped samples of this species.  In addition to calculations for CaF, we also report experimental tests in atomic helium of one of the key requirements, the production of bichromatic forces in a multi-level system with several simultaneously cycling transitions, not all of which contribute to the force.

\begin{figure}
\includegraphics[width=\columnwidth]{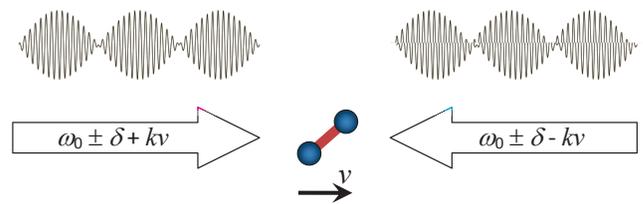}
\caption{(color online). Simplified sketch of the optical bichromatic force on a diatomic molecule.  Two-color beams impinge from each direction, offset by frequencies $\pm \delta$ from resonance at $\omega_0$. For large molecular velocities $v$, Doppler offsets of $\pm kv$ must also be added.  The frequency pairs at $\pm \delta$ give rise to a series of rf beat notes, each with a pulse area of approximately $\pi$.  Alternating cycles of excitation from the right and stimulated emission from the left produce a large decelerating force directed to the left.  The direction is controlled by the relative phase of the beat notes from the right and the left.  }
\label{fig:BCF_concept}
\end{figure}

Of the several configurations utilizing two-frequency beams that can yield rectified optical forces \cite{Cashen03,Grimm96},  it is well-established that the counterpropagating-beam BCF configuration shown in Fig. 1 works best for deceleration and cooling of a two-level system \cite{Grimm90,Grimm94,Grimm96,Soding97,Cashen01,Cashen02,Cashen03,Partlow04,Partlow04b}.  Here a pair of oppositely directed two-color beams is configured so that in the frame of a moving atom or molecule, the Doppler-shifted frequencies in each beam are $\omega_0 \pm \delta$, thus creating a pair of counterpropagating beat note pulse trains.  If the laser power is chosen so that the area of each rf beat-note pulse is approximately $\pi$, the atom or molecule will experience a sequence of alternating excitations and stimulated de-excitations. A large deceleration results if the excitations come from the right-hand beam and the de-excitations from the left.  If the beat notes have a relative rf phase of $\pi/2$, heuristic arguments suggest that the correct sequence occurs on average 3/4 of the time, while the incorrect sequence, leading to acceleration, occurs 1/4 of the time.  Numerical calculations and experiments support this argument.  If the cycling rate is fast compared to the radiative decay time, the average force can greatly exceed the ordinary radiative force.

This concept was pioneered both experimentally and theoretically in the late 1980's and 1990's, largely by the group of Yatsenko \cite{Voitsekhovich89,Voitsekhovich94,Yatsenko04} and by Grimm and coworkers \cite{Grimm90,Grimm94,Grimm96}, who ultimately slowed a cesium atomic beam to 8 m/s \cite{Soding97}.  More recently the Metcalf group at Stony Brook has extensively investigated the BCF on metastable atomic helium (He*) beams.  They have demonstrated both transverse collimation and longitudinal slowing by 325 m/s in a region just a few centimeters long \cite{Cashen03,Cashen01,Cashen02,Partlow04,Partlow04b} and devised improved theoretical treatments.  However, BCFs have not so far been used to produce ultracold atoms for magnetic or optical trapping.

In view of this long and mostly successful history, it seems odd that BCF decelerators have found only limited use.  In the case of alkali atoms, this is largely because direct or indirect vapor-cell loading often eliminates the need for a decelerator, and it is not particularly difficult to construct a Zeeman slower when one is needed.  Another consideration is that it was technically difficult until recently to produce high-power cw laser beams at multiple frequencies with good control over the relative rf phase.  There appears to be just one previous attempt to apply a BCF-like force to a molecule, a 1994 experiment in which counterpropagating mode-locked pulse trains produced a small transverse deflection in a Na$_2$ beam \cite{Voitsekhovich94}, with a total momentum increment up to $20 \hbar k$ that was limited by unwanted optical pumping into dark states.  In that work, the time sequencing of counterpropagating pulse pairs played a similar role to the phase shift between beat notes in the counterpropagating-beam BCF of Fig. \ref{fig:BCF_concept}.

Assuming that technical problems such as optical pumping can be overcome, the force multiplication offered by the BCF at large detunings $\delta$ is even more attractive for near-cycling molecular transitions than for atoms, because the available interaction time is generally constrained by out-of-system radiative decays.  In Section \ref{sec:BCF} of this paper, we discuss general considerations governing BCFs in a molecular transition, and in Section \ref{sec:CaF} we consider the specific example of CaF, still with an emphasis on readily generalizable approximations.  In Section \ref{sec:He} we present a simple experimental test of the BCF in a multilevel atomic system, metastable He* atoms illuminated by $\pi$-polarized light.  These experiments test both the feasibility of using the BCF in a multilevel system and the sensitivity of the force to energy level shifts away from the laser center frequency $\omega_0$.

\section{Basic properties of the BCF for a near-cycling transition\label{sec:BCF}}
In this section we begin with a summary of the basic BCF parameters for a two-level system, then discuss the modifications necessary for a near-cycling transition that may also contain multiple internal levels.  For the two-level case, both the heuristic argument given above and theoretical dressed-atom treatments \cite{Grimm96,Cashen02,Cashen03,Partlow04} show that the maximum attainable bichromatic force for a two-level atom is given approximately by
\begin{eqnarray}
F_b^{\text{TLA}}=\hbar k \delta/\pi
\label{eq:FbTLA}.
\end{eqnarray}

This is larger than the radiative force $F_{\text{rad}}^{\text{TLA}}=\hbar k \gamma/2$ by a factor of $\simeq 0.64~\delta/\gamma$.  The optimal Rabi frequency for each beam is $\Omega = \sqrt{3/2}\,\delta$, slightly larger than the $\pi$-pulse condition of $\Omega = (\pi/4)\delta$.  The corresponding laser irradiance for each beam is given by
\begin{eqnarray}
I_{b}=3\left(\frac{\delta}{\gamma}\right)^2 I_s\label{eq:Irrad},
\end{eqnarray}
where $I_s$ is the ordinary saturation irradiance at resonance \cite{Metcalf99} and $\gamma$ is the upper-level decay rate.  Numerical calculations and experiments also show that the BCF has a much wider velocity range than the radiative force,
\begin{eqnarray}
\Delta v_b \simeq \delta/k\label{eq:range},
\end{eqnarray}
 compared with $\Delta v_{\text{rad}} \simeq \gamma/k$ for a two-level system.

Clearly it is advantageous to use large detunings $\delta$, but a practical limit is set by the quadratic scaling of the required laser power in Eq. \ref{eq:Irrad}, which will eventually outstrip any available laser resources.  In actual practice detunings of $\delta \simeq 200 \gamma$ have been used successfully for metastable He atoms, although at $300 \gamma$ we have seen evidence that the force begins to lose the symmetric velocity profile characteristic of rectified optical forces \cite{Chieda11b}.

Another important property of the BCF is that it usually has a rather sharp lower velocity limit, which gives rise to cooling of the velocity distribution in addition to slowing, because the slowed molecules will ``pile up'' at the lower velocity limit.  This is apparent, for example, in the results reported in Ref. \cite{Soding97}.  BCF cooling is the primary topic of a recent discussion by Metcalf \cite{Metcalf08}, who also mentions briefly the possibility of applying counterpropagating-beam BCFs to molecules.

We now turn to the case of a near-cycling transition, initially without considering its degeneracy.  A typical example is the $A-X$, ($0-0$) band in MgF, CaF, or SrF, for which the Franck-Condon factors are in the range $0.98-0.99$ \cite{Wall08,Pelegrini05,Shuman09}, so the molecules will be lost to out-of-system decay in about $50-100$ spontaneous decay cycles unless additional repump lasers are added.  In this case the above relations are still valid, but the maximum BCF momentum transfer will generally be limited by the time $T_{\text{loss}}$ until an out-of-system decay occurs.  To determine $T_{\text{loss}}$ for a molecule with spontaneous decay lifetime $1/\gamma$ and Franck-Condon factor $\mathfrak{F}$, an estimate is needed for the probability $P_e$ that the molecule is in the excited state.  For a single laser beam at the optimal irradiance $I_b$, this would be $P_e = 0.3$, but for a full four-beam BCF configuration the situation is more complicated.  The bichromatic beat-note phase shift of $\pi/2$ corresponds to the condition that the molecules follow dressed-atom trajectories comprised primarily of the ground-state wave function \cite{Soding97,Cashen02}, but the exact value of $P_e$ will depend on averaging over velocity profiles and laser mode profiles.  Here we will assume that $P_e \approx 1/3$, near the center of the possible range, and the loss-limited interaction time is given by
\begin{eqnarray}
T_{\text{loss}} = \frac{ 1}{\gamma P_e(1-\mathfrak{F} )}.
\label{eq:T_loss}
\end{eqnarray}

The maximum possible velocity decrease from the BCF is given by the smaller of $\Delta v_b$ from Eq. \ref{eq:range} and the value obtained if deceleration continues for the full available time $T_{\text{loss}}$,
\begin{eqnarray}
\Delta v_{\text{loss}} = T_{\text{loss}} \frac{F_b}{M},
\label{eq:dv_loss}
\end{eqnarray}
where $M$ is the mass of the molecule.  In the two-level case this simplifies to
\begin{eqnarray}
\Delta v_{\text{loss}}^{\text{TLA}} \simeq \frac{3}{\gamma (1 - \mathfrak{F})} \frac{\hbar k \delta}{\pi M}.
\end{eqnarray}

\begin{figure}
\includegraphics[width=0.8\columnwidth]{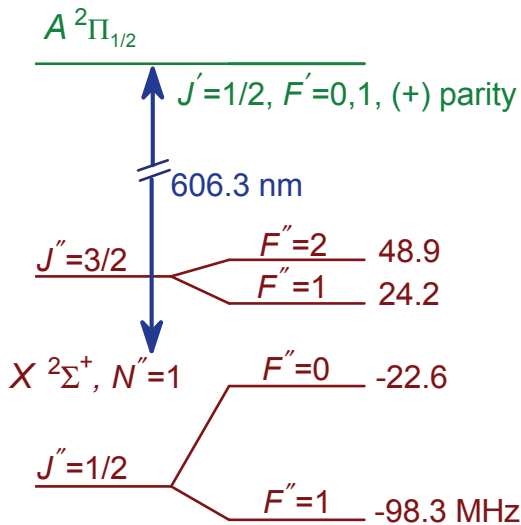}
\caption{(color online). Rotationally closed transition from the $N''$=1 level of the $X\,^2\Sigma^+$, $v''$=0 ground state of CaF to $J'$=1/2 in the $A\,^2\Pi_{1/2}$, $v'$=0 state.  The corresponding transition in SrF was used for laser cooling in Ref. \cite{Shuman10}.  The ground-state fine structure and hyperfine structure give rise to a total of 12 $m_{F'}$ and $m_{F''}$ levels, not shown.  In traditional notation \cite{Field75} this is a combination of the near-degenerate $Q_{12}$(0.5) and $P_{11}$(1.5) branches.}
\label{fig:P_branch}
\end{figure}

\subsection{Degeneracy and near-degeneracy\label{sec:degeneracies}}
In any real molecular transition, the effects of small internal splittings and degeneracies must also be considered, including fine structure (fs) and hyperfine structure (hfs) as well as the magnetic quantum numbers $m_J$ or $m_F$.  For example, the CaF and SrF molecules each have nuclear and electronic spins $I$=$S$=1/2, leading to structure such as that shown in Fig. \ref{fig:P_branch} for CaF.  The analogous rotationally closed system in SrF was used in recent demonstrations of Doppler and Sisyphus cooling \cite{Shuman09,Shuman10}.   In either species, the upper-state hfs is negligibly small, but still gives rise to four near-degenerate $m_{F'}$ sublevels, and the $N''$~=~1 lower state has four well-separated hfs levels with a total of 12 $m_F$ sublevels.  To achieve Doppler and Sisyphus cooling \cite{Shuman10}, the states are intentionally mixed using a combination of Zeeman mixing and several repumping frequencies that are introduced using electro-optical modulators.  A vibrational repumping laser is also used to extend the value of $T_{\text{loss}}$.   The net effect is to introduce a degeneracy that reduces the two-level radiative force, because the molecules spend only about 1/7 of their time in the less numerous sublevels of the excited state \cite{Shuman09}.

There are two important differences between the BCF case and that of Doppler cooling:  First, the large Rabi frequency used for the BCF causes power broadening that can make many or all of the fs and hfs levels nearly resonant, reducing the need for added repumping frequencies.  Second, a new problem arises because the transitions between the various sublevels have varying rates that can easily span an order of magnitude \cite{Wall08}, and some will be detuned from exact resonance, so not all of them can simultaneously be driven at the optimal rate for the BCF.

\begin{figure}
\includegraphics[width=\columnwidth]{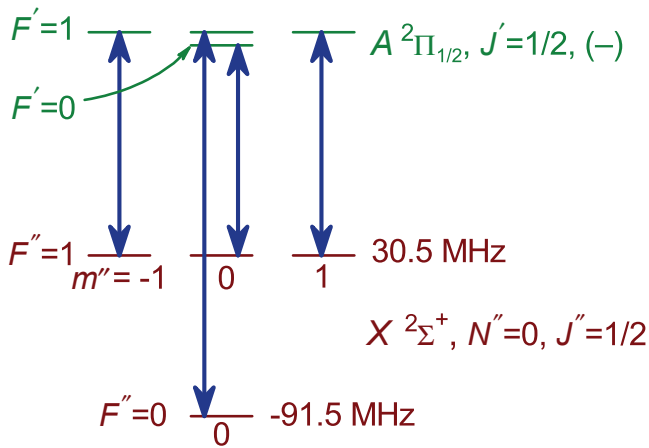}
\caption{(color online).  Arrows show the four quasi-cycling transitions between the $m_{F''}$ and $m_{F'}$ sublevels of the $Q_{11}$(0.5) branch in CaF, when illuminated with $\pi$-polarized light.  All four line strengths are the same.  Energies are not drawn to scale. Radiative decay to $N''=2$ is allowed, so a rotational repumping laser tuned 62 GHz to the red is required.}
\label{fig:Q_branch}
\end{figure}

To account for these effects, the results of the preceding section must be modified.  The excited-state probability $P_e$, previously taken to be $\approx 1/3$, is estimated using a weighted degeneracy factor that has the same two-level limit,
\begin{eqnarray}
P_e \approx \frac{g_e/2}{g_e/2 + g_a + g_d}.
\label{eq:P_e}
\end{eqnarray}
Here $g_e$ is the degeneracy of the excited state, $g_a$ is the degeneracy of the ground-state levels that are active in the BCF, and $g_d$ is the degeneracy of ``dark'' ground state levels that are populated but do not participate in the BCF.  There is a certain arbitrariness to the method used for adjusting the excited-state fraction to a slightly non-statistical value, since for a strongly coupled multilevel system, the actual value is determined by system-dependent coherent cycling.  We use the $g_e/2$ weighting because it correctly treats the effects of weakly-coupled dark levels on a system with multiple decoupled two-level cycles, such as the scheme shown in Fig. \ref{fig:Q_branch}.  Repumping of these dark levels is discussed in Sec. \ref{sec:repumping}.

For the bichromatic force, the level-averaged cycling rate is reduced if only a fraction of the ground-state levels are active, again requiring a weighted degeneracy factor.  An additional ``force reduction factor'' $\eta$ is introduced to account for possible further reductions in the BCF due to variations in line strength and frequency shifts from level to level,
\begin{eqnarray}
F_b = \frac{\hbar k \delta}{\pi} \frac{g_e/2 + g_a}{g_e/2 + g_a + g_d} \, \eta.
\label{eq:Fb}
\end{eqnarray}
The degeneracy factor is again chosen to describe a system with multiple decoupled two-level cycles, and system-dependent deviations due to interference in multilevel coherent cycling are taken to be part of $\eta$. The loss time is still given by Eq. \ref{eq:T_loss}, and the velocity change $\delta v$ by the smaller of Eqs. \ref{eq:range} and \ref{eq:dv_loss}, providing that the revised values of $P_e$ and $F_b$ are used.  The ordinary radiative force is also reduced, although in this case the impact of degeneracies and dark states is purely statistical:
\begin{eqnarray}
F_{\text{rad}} = \frac{\hbar k \gamma}{2} \frac{g_e}{g_e + g_a + g_d}.
\label{eq:Frad}
\end{eqnarray}

To deal with the negative impacts of degeneracies and inexact Rabi frequencies, at least three experimental schemes are possible, any of which might be preferable depending on the particular level structure and experimental objectives for a specific molecule:

(1) Ignore the degeneracy, apart from adding a small magnetic field to circumvent optical pumping into coherent dark states, as in Ref. \cite{Shuman09}.  A significant average BCF is still be possible, because as long as the bichromatic beat notes are phased properly the resulting force is always positive or zero, never negative, even for grossly incorrect values of the laser frequencies and the Rabi frequency.  Assuming the force is applied for many radiative cycles, each molecule will spend some of its time in the sublevels for which the BCF is near-optimal, and every molecule will see the same time-averaged force.  A complete quantitative model would require extending the numerical methods of Refs. \cite{Grimm96,Soding97,Partlow04} to a multi-level set of optical Bloch equations (OBEs) tailored to each specific system.  However, two-level OBE simulations are sufficient to establish the tolerance of the BCF to non-ideal configurations.  The results of such calculations for non-optimal Rabi frequencies are shown in Fig. 1 of Ref. \cite{Soding97} and Fig. 62. of Ref. \cite{Cashen02}, indicating that substantial forces are still produced for variations of up to about 20\%.  We have used similar numerical modeling to investigate the effects of frequency shifts away from the nominal resonant frequency $\omega_0$, an issue that has not previously been studied.  Our modeling shows that as the shift increases, the BCF force profile always retains about the same velocity range, but the average force is reduced.  Table \ref{tbl:EtaVals} shows the calculated force reduction factor $\eta$ as a function of the shift, measured as a fraction of the bichromatic detuning $\delta$.  In Section \ref{sec:He} we describe an experimental test in metastable atomic helium.

\begin{table}
\caption{\label{tbl:EtaVals}Calculated force reduction factors $\eta$ for Eq. \ref{eq:Fb}, for levels shifted from the laser center frequency $\omega_0$.  Shifts are tabulated as a fraction of the bichromatic detuning $\delta$.}
\begin{ruledtabular}
\begin{tabular}{cc}
Fractional Shift &	$\eta$\\
\hline
0 & 1.0\\
0.1 & 0.76\\
0.2 & 0.45\\
0.3 & 0.25\\
\end{tabular}
\end{ruledtabular}
\end{table}

(2) Use brief pulses of the BCF beams with $\sigma^-$ circular polarization, alternating with optical pumping using $\sigma^+$ polarization.  If the fine and hyperfine structure is not too large, the optical pumping can include all of the fs and hfs levels by use of modulator-induced sidebands as in Ref. \cite{Shuman09}.  In this case the population will mainly cycle between extremal $m_F$ levels, creating a temporary two-level system that is destroyed after several radiative decay cycles.  Depending on the degeneracies and decay rates, the BCF beam could be left on for at least a few radiative decay times, followed by optical pumping for perhaps 4-10 decay times.  Taking this repumping interval into account as a loss in the time-averaged force, values of $\eta$ in the range of about 0.2-0.6 might be expected.

(3) Circumvent the problem by locating a rotational branch for which the upper and lower levels have similar sublevels and no dark states, such as the $Q_{11}(0.5)$ branch in CaF shown in Fig. \ref{fig:Q_branch}.  Such a transition will generally not be rotationally closed (for example, decay to $N''$~=~2 can occur in this example), but a rotational repumping laser can plug this hole.  In such a case there will generally be several sets of levels that cycle independently in the BCF laser field, although the molecules will randomly switch from one cycle to another when spontaneous decay occurs.  This occasional redistribution of $m_F$ should make little difference to the BCF, and helps to insure that each molecule sees the same time-averaged force. In the example of Fig. \ref{fig:Q_branch} the upper and lower states, each with $J$=1/2, have $F$=0 and $F$=1 components with no fine structure.  For $\pi$-polarized light there are four separate pairs of cycling magnetic sublevels, all with the same line strength except for the effects of frequency shifts due to the hyperfine splitting.  Here the force reduction factor $\eta$ is determined mainly by these frequency shifts, and can be estimated from Table \ref{tbl:EtaVals}.  There will also be degeneracy factors due to the level multiplicities $g_a$ and $g_d$, as in method (1), depending on the level structure and the repumping scheme.

\subsection{Repumping\label{sec:repumping}}

Although the BCF can serve as a highly effective force multiplier, in most cases it will still be necessary to use repumping lasers to avoid premature loss of the molecules to ``dark" states before the deceleration and cooling process is complete.  Compared to repumping in conventional radiative laser cooling schemes, the rapid coherent cycling of the BCF transition introduces a few new issues, but also a few new possibilities.  Assuming that a weak cw laser is used to repump a dark level to the excited state, the situation is a classic strong-pump, weak-probe scenario, and the repumping rates can be estimated using ordinary incoherent rate equations.

If the repumping laser were on exact resonance to the excited state of the cycling transition, it would be sufficient to use a laser irradiance slightly above the saturation irradiance $I_s$ for the cycling transition, typically just a few tens of mW/cm$^2$.  This holds true regardless of the strength of the dark-state transition, because if the decay branching fraction to a particular state is $\alpha$, the square of the dipole matrix element for the repumping transition scales as 1/$\alpha$, but the required repumping rate needs only to keep up with the loss rate, introducing a compensating factor of $\alpha$ in the required irradiance.

For the BCF case, the situation is more complex because of the time-dependent interference of the counterpropagating bichromatic beams, which causes the dressed-atom levels to vary in energy on a distance scale of about $\lambda/4$ \cite{Cashen02, Yatsenko04}. For a molecular beam velocity of $\simeq$~100~m/s, this causes time-dependent shifts of the dressed energies on a nanosecond time scale, with a complicated structure that varies inhomogeneously and also evolves with the deceleration of each atom, due to varying Doppler shifts.  Although an exact solution would require very detailed numerical modeling of the multilevel optical Bloch equations, a much easier approach is to consider the incoherent excitation rate to a state whose effective linewidth is given by the energy range spanned by the dressed energy levels.  This is a good approximation so long as the time scale of the energy shifts is much faster than the required repumping rates.  The resulting effective linewidth is slightly less than $\delta$, depending on the BCF Rabi frequency, as can be seen by examining the dressed energy levels in Fig. 6.4 of Ref. \cite{Cashen02} or Fig. 3 of Ref. \cite{Yatsenko04}.  It should be noted that when the diabatic dressed-state energies change by more than $\delta$ on these plots, this corresponds to emission or absorption of a photon, while the actual atomic energy remains within the range $(-\delta /2, \delta /2)$.

This effective linewidth is actually rather convenient because it is typically sufficient to span all of the fs and hfs sublevels of a given rovibrational transition, making it unnecessary to use modulators to produce multiple repumping frequencies tuned to each sublevel.  On the other hand, the irradiance required for repumping scales with the linewidth, and thus is increased from $I_s$ by a factor of approximately $\delta / \gamma$.  Because this scaling is only linear with $\delta$ rather than quadratic as for the main BCF beams, the laser requirements will typically still be modest, on the scale of several hundred mW/cm$^2$, requiring only several mW to produce a mm-size beam.

Up to this point, we have assumed that all of the ground-state quantum levels, both active and dark, are statistically equilibrated by incoherent radiation and repumping.  This is actually something of a worst-case assumption, since the coherently cycling dressed-atom levels offer opportunities for more subtle manipulation of the populations.  A well-chosen repumping frequency might in principle be used to select a dressed excited state that has little pre-existing population, although this particular method for one-way repumping would be likely to wash out when averaging over atoms.  More interesting is the prospect of a coherent repumping scheme that is synchronized to the bichromatic beat note frequency, either by the use of pulse trains with rf repetition frequencies or multi-color excitation.  Such a scheme might selectively populate dressed-atom states with a consistent phase, enhancing the BCF while producing a non-equilibrium population distribution.  A systematic investigation of such schemes will require very careful nanosecond-scale modeling of the full OBEs, which we hope can be conducted in the future.

\section{Application to calcium monofluoride\label{sec:CaF}}

The CaF molecule makes a good test case for several reasons.  As noted in the Introduction, its spectrum has been extensively studied.  Its large molecule-fixed dipole moment of 3.07 Debye  \cite{Childs84} and its small rotational constant  of 0.3437 cm$^{-1}$ \cite{Childs81} make it attractive for studies of dipolar interactions in applied external fields, although like many polar molecules, CaF has a  $\Lambda$=0 ground state and thus has a linear Stark effect only when the external field is large enough to induce rotational mixing.  CaF has recently been used to demonstrate a buffer-gas cooled beam \cite{Maussang05} and a scheme for alternate-gradient focusing  \cite{Wall09}.   The Rydberg states of CaF are of special interest because of the effects of the very large dipole moment of the CaF$^+$ core, 8.9~D relative to the center of mass \cite{Gittins01}.  Finally, its accessible ionization energy (IE) of 46796.4 cm$^{-1}$ facilitates resonant multiphoton ionization (REMPI) and the excitation of high Rydberg states.

Most important, CaF is also an excellent candidate for tests of BCF slowing and cooling.   Relative to SrF, the CaF molecule has about half the mass, making it easier to slow, and its fine structure is smaller \cite{Childs81}, making it easier to achieve a bichromatic detuning $\delta$ larger than the fs and hfs level spacings.  Either of the two near-cycling $A-X$, (0-0) transitions introduced in Figs. \ref{fig:P_branch} and \ref{fig:Q_branch} could be used.  Here we consider the example of Fig. \ref{fig:Q_branch}, the $Q_{11}(0.5)$ branch at 606.3 nm, which has four essentially independent stimulated cycling transitions and corresponds to scheme (3) above.  Because 1/3 of the spontaneous decays occur to $N''$~=~2, a rotational repumping laser is necessary.  Vibrationally the (0-0) band has a Franck-Condon factor of about 0.987 \cite{Wall08,Pelegrini05} that is even higher than for the analogous $A-X$ system of SrF.  Nevertheless, for very large momentum transfers a vibrational repumping laser will be necessary on the (0-1) band at 628.5 nm, which should reduce the loss rate by at least a factor of 50 \cite{Shuman10}.  No complications are expected from coherences between the four separate cycles in Fig. \ref{fig:Q_branch}, because when spontaneous decay events occasionally transfer molecules from one cycle to another, they randomly reset the phase.  Thus the sample is effectively divided into four sub-samples that each experience a BCF, although it will be slightly smaller for $F''$=0 because of the 122 MHz hfs shift relative to $F''$=1.  In Section \ref{sec:He} we experimentally evaluate a similar level scheme using an atomic helium beam.

A well-collimated and relatively slow beam source will be needed if the BCF is to provide deceleration sufficient for loading a cold molecule trap.  Fortunately, intense and rotationally cold beams are possible with laser ablation sources, and laser-induced fluorescence has been used successfully by several groups to detect and characterize molecular beams of CaF.  At ordinary thermal temperatures, an intense supersonic beam can be formed by methods such as that of Ref. \cite{Wall08}, in which a pulsed jet of argon and SF$_6$ is placed close to a calcium laser ablation target, yielding a pulsed jet with a rotational temperature of just 3 K.  To obtain lower common-mode velocities, the buffer-gas cooling method has been employed successfully both for CaF \cite{Maussang05} and SrF \cite{DeMille11b}, and can provide bright beams with velocities of about 140 m/s.  If a bichromatic detuning of $\delta/2 \pi$=250~MHz is used, Eq. \ref{eq:range} predicts a velocity range $\Delta v_b$ = 150 m/s that would permit slowing all the way to zero velocity.

In table \ref{CaFparms} we list representative parameters based on recent measurements of the $A$ state lifetime and the $Q_{11}(0.5)$ saturation irradiance ($I_s$ = 22.2~mW/cm$^2$) obtained using laser-induced fluorescence \cite{Wall08}.  In keeping with the above discussion, we assume a bichromatic detuning of 250 MHz, or $\delta$=30.2~$\gamma$.  The excited-state probability $P_e$ is taken from Eq. \ref{eq:P_e} using $g_e$=4, $g_a$=4, and $g_d$=13, based on the 13 dark sublevels of the $N''$=2 state.  The force reduction factor $\eta$ due to repumping is taken to be 0.78, based on an optimal BCF for the three $F''$=1 levels and a very inefficient BCF for the detuned $F''$=0 level.  The estimated laser irradiance requirement of 60~W/cm$^2$ for each of the four BCF beam components (two beams, each with two colors) is well within the range that can be obtained with a single cw dye laser using a beam diameter of about 0.5 mm.  A factor of two is gained because the counterpropagating beam can be obtained by simple reflection of the copropagating beam on a mirror spaced 7.5 cm from the molecular beam.  At this spacing, the optical propagation delay produces the required $\pi/2$ phase shift between the bichromatic beat notes in the forward and reverse-directed beams.  It would also be possible to work with larger beams or larger bichromatic detunings by using multiple reflections between a pair of off-center mirrors separated by 30 cm, which would again maintain the bichromatic beat note phasing.

\begin{table}
\caption{\label{CaFparms}Representative BCF parameters for CaF, assuming no vibrational repumping, $P_e$ = 2/19, $\eta$ = 0.78, and a degeneracy factor of 6/19 in Eq. \ref{eq:Fb}.  With vibrational repumping, $T_{\text{loss}}$ would increase to $\approx$~350~$\mu$s and $\Delta v_\textrm{loss}$ would exceed $\Delta v_b$, although the forces $F_b$ and $F_{\text{rad}}$ would be reduced by nearly a factor of two due to the additional degeneracy of 17 dark states with $v''$=1.}
\begin{ruledtabular}
\begin{tabular}{lll}
Parameter &	Symbol &Value\\
\hline
Bichromatic detuning	& $\delta/2 \pi$ & 250 MHz\\
Deceleration & $a$ & 1.4 $\times$ $10^6$ m/s$^2$\\
Bichromatic velocity range	& $\Delta v_b$ & 150 m/s\\
Loss time & $T_{\text{loss}}$ & 14 $\mu$s \\
Loss-limited velocity range	& $\Delta v_{\text{loss}}$ & 19.4 m/s\\
Optimal irradiance & $I_b$ & 60 W/cm$^2$\\
Ratio of BCF to rad. force & $F_b$ : $F_{\text{rad}}$ & 12.4\\
\end{tabular}
\end{ruledtabular}
\end{table}

In this example the velocity change is constrained to just 19.4 m/s due to vibrational loss to $v''$=1.  However, the introduction of a vibrational repumping laser would remedy this by increasing $\Delta v_{\text{loss}}$ to 250 m/s, larger than the full bichromatic velocity range $\Delta v_b$.  This would also allow for significant longitudinal cooling if the velocity range were chosen with a lower limit at a small non-zero value, since there would be time enough for atoms to accumulate at the lower velocity limit of the range.

There is an additional loss mechanism that has so far not been taken into account, the possibility of off-resonant excitation into the wrong rotational branch caused by the intense BCF beams.  In this example the largest such contribution arises for molecules that have radiatively decayed to the $N''$=2, level, which can be excited on the $R_{12}(3/2)$ branch by the BCF beams before they are repumped to $N''$=1.  This transition is off-resonant by 20.4 GHz, or about 2500~$\gamma$.  Because this greatly exceeds the Rabi frequency, the estimated transition rate is small, roughly 3000/s based on a simple rate-equation estimate, and still allows for the full bichromatic slowing time $T_b$.  However, in other molecules this would not always be the case, and losses to off-resonant excitation must always be considered carefully as a possible limiting factor in BCF deceleration and cooling.

Another effect of off-resonant transitions is that they lead to ac Stark shifts that vary from level to level, shifting the transitions used for the BCF from their optimal values.  Because the BCF is robust against small frequency shifts this should be insignificant in most cases --- for CaF, the estimated shifts are just a few MHz.  A particularly stringent test occurs in He*, where the $2\,P_2$ level is close enough to the $2\,P_1$ level to cause large shifts, as we discuss in Section \ref{sec:He}.

So far we have considered only longitudinal slowing and cooling.  In some cases, only transverse deflection or focusing is needed, and this is a far easier to achieve.  For transverse deflection, an interaction length of 5 mm along a supersonic CaF beam gives an interaction time of 10 $\mu$s at 500 m/s, somewhat less than the estimated time $T_{\text{loss}}$ for radiative loss to (0-1) transitions if there is no vibrational repumping laser.  The resulting transverse velocity increment of 14 m/s corresponds to a deflection angle of 28 mrad, more than adequate to achieve separation from the original beam.  The momentum transfer of about 1200 $\hbar k$ would considerably exceed all previous results for molecular deflection by optical forces: 1 $\hbar k$ and 20 $\hbar k$ for Na$_2$ \cite{Voitsekhovich94,Herrmann79}, and most recently, 150~$\hbar k$ for SrF \cite{Shuman09}.

An interesting variation on the transverse deflector would be to use transform-limited microsecond or nanosecond pulses for a ``one-shot'' deflector, operating for only a single radiative lifetime so that no cycling transition is needed.  This could be accomplished using very large detunings and high pulse irradiances, which are limited only by the onset of ionization of the molecules.  Based on recent atomic helium studies, detunings up to at least 200~$\gamma$ (1.65 GHz for CaF) could be expected to produce useful BCFs.  This scheme could provide state-selective deflection for nearly any molecule with accessible transitions.

\section{Experimental tests in atomic helium\label{sec:He}}
\subsection{Atomic helium as an experimental analog to CaF\label{sec:He-CaF}}

\begin{figure}
\includegraphics[width=\columnwidth]{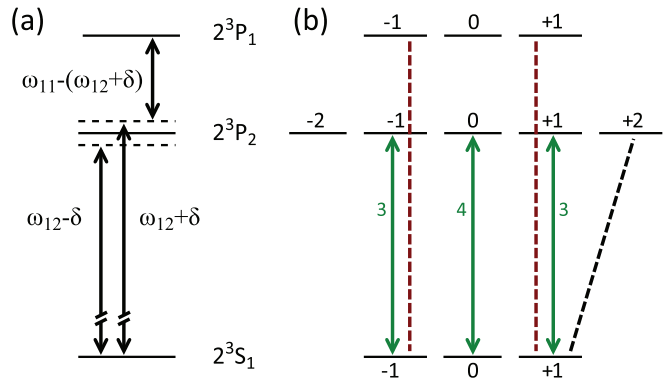}%
\caption{(color online).  (a) Relevant energy levels in metastable helium, showing the two BCF frequency components and an off-resonant coupling to the $2\,^3P_1$ state, discussed in the text. The distant $J' = 0$ fine-structure level plays no significant role and is not shown.  (b) Zeeman sublevels, with vertical solid arrows indicating allowed transitions for $\pi$-polarized light, labeled with their relative transition strengths \cite{Metcalf99}.  The vertical dashed lines indicate transitions giving rise to off-resonant excitation and ac Stark shifts, and the angled dashed line indicates is the transition excited by $\sigma^+$ polarization in the conventional BCF configuation for He*.}
\label{fig:helevels}
\end{figure}

As discussed in Sections \ref{sec:BCF} and \ref{sec:CaF}, one major departure of the BCF for a realistic molecular transition from the two-level atomic model is the simultaneous cycling of several $m_J$ or $m_F$ levels. This occurs both in schemes (1) and (3) of Section \ref{sec:BCF}.  Here we experimentally evaluate this scenario using the 1083 nm $2\,^3S_1 \leftrightarrow 2\,^3P_2$ transition in metastable helium (He*).  The usual BCF scheme for He* cycles the (1,2) component of this transition by the use of $\sigma^+$ circularly-polarized light~\cite{Cashen01,Cashen03}, where the notation (1,2) is used to denote $m$=1 $\leftrightarrow$ $m'$=2.  The usual arrangement has the advantages of providing the largest possible transition strength as well as providing a near-perfect two-level atom due to the absence of hyperfine structure. If instead the bichromatic slowing beams are linearly polarized ($\pi$ polarization), then the ($-1$,$-1$), (0,0) and (1,1) components will all be driven simultaneously as shown in Fig.~\ref{fig:helevels}. This configuration is rather similar to the $Q_{11}(0.5)$ transition in CaF, shown in Fig.~\ref{fig:Q_branch}.  Although the overall $2\,^3S_1 \leftrightarrow 2\,^3P_{J'}$ system is closed, an added complication in helium is that there is strong off-resonant coupling to the $2\,^3P_1$ fine-structure level, providing a good opportunity to test the effects of an off-resonant level in a situation where it does not lead directly to permanent population loss.  Here we study this configuration using a very large bichromatic detuning of $\delta/2 \pi = 300$~MHz, corresponding to nearly 200 natural linewidths.

The optical power requirements inevitably increase when moving from a two-level to a multi-level scheme.  Here we simultaneously drive three transitions, two optimally and one slightly more strongly than optimally. Specifically, the irradiance is chosen such that the optimal Rabi frequency criterion $\Omega_{R}=\sqrt{(3/2)}\,\delta$~\cite{Cashen01} holds for the (-1,-1) and (1,1) components.  This requires exactly twice the irradiance needed to drive the (1,2) component using $\sigma^+$-polarized light, or about 68~W/cm$^2$ for $\delta/2 \pi$~=~300~MHz.  The (0,0) transition will then experience $\Omega_{R}= \sqrt{2} \delta$ in the center of BCF laser beams, too high by about 15\%.  In practice, the BCF must be averaged over the Gaussian laser beam profile.  Based on measurements of the laser power dependence when using $\sigma^+$ polarization, we estimate that the spatially averaged BCF is reduced only by about 5\% when the peak irradiance is too high by 15\%.

As already mentioned, another important consideration is the effect of off-resonant coupling to the $2\,^{3}P_{1}$ fine-structure level, which lies 2.29~GHz above the target $2\,^3P_2$ state. While the (0,0) component of this transition is forbidden, the (1,1) and (-1,-1) transitions are allowed, as denoted by the vertical dashed arrows in Fig~\ref{fig:helevels}, with the same line strengths as for the equivalent componetns of the $2\,^3S_1 - 2\,^3P_1$ transition.  This does not cause loss, since $2\,^3P_1$ decays radiatively back to $2\,^3S_1$, but it can cause two other issues:

(1) If the laser power is sufficiently high, the off-resonant transition rate can be significant.  This rate is still much slower than the cycling rate for the target transition, and it should be incoherent, so it is not expected to directly interfere with the BCF.  However, if it is comparable to the spontaneous decay rate $\gamma = 10^7$/s, it will result in substantial populations in the two coupled sublevels $m' = \pm1$ of the $2\,^3P_1$ state, which do not participate in the BCF.  In the present case we estimate an excitation rate of $0.027 \gamma$, too small to cause any observable force reduction.

(2) There can be significant level-dependent ac Stark shifts, in this case for the $2\,^3S_1, m = \pm 1$ levels, which shift the bichromatic detunings from $\pm \delta$ into an asymmetric configuration.  These ac Stark shifts can be written as
\begin{equation}
\Delta \omega_{\text{ac}} = \frac{\Omega_{m,m'}}{2 \Delta^2},
\end{equation}
where $\Omega_{m,m'}$ is the Rabi frequency for the ($m,m'$) component of the $2\,^3S_1 - 2\,^3P_2$ transition, and $\Delta$ is the detuning from $2\,^3P_1$, approximately $\Delta = (2.29$~GHz)$\times 2 \pi$ ~$\pm$~$\delta$ for the BCF frequency components at $\mp \delta$.  The Stark shifts are time-dependent on a nanosecond scale due to the interference of the bichromatic beams, much as for the repumping of molecular transitions discussed in Section \ref{sec:repumping}.  The time-averaged shift is $\Delta \omega_{\text{ac}}/2 \pi = -120$~MHz for $m = \pm 1$.  This is more than 1/3 of the bichromatic detuning $\delta$, and using Table \ref{tbl:EtaVals}, it leads to a force reduction factor of $\eta \simeq 0.2$.  However, the $m = 0$ sublevel is nearly unaffected, since it is shifted only by interaction with the distant $2\,^3P_0$ level, detuned by $\Delta/2 \pi \approx$~32 GHz.

Taking all of these effects into account and averaging over the three $m$ sublevels of the $2\,^3S_1$ state, we estimate that that the effective bichromatic force should be $F_{\text{eff}} \approx 0.45$~$F_b$, where $F_b$ is the ordinary BCF for $\sigma^+$ polarization.  While this loss of force is significant, the interaction time of the atoms with the BCF beams is about 26~$\mu$s, about 5.6 times the bichromatic slowing time $T_b$ for $\sigma^+$ light.  This is sufficient to slow the atoms through the full BCF velocity range even with the reduced force, so we predict little effect on the observed velocity profiles.

\subsection{Experimental design\label{exp_des}}

\begin{figure*}
\includegraphics[width=0.8\textwidth]{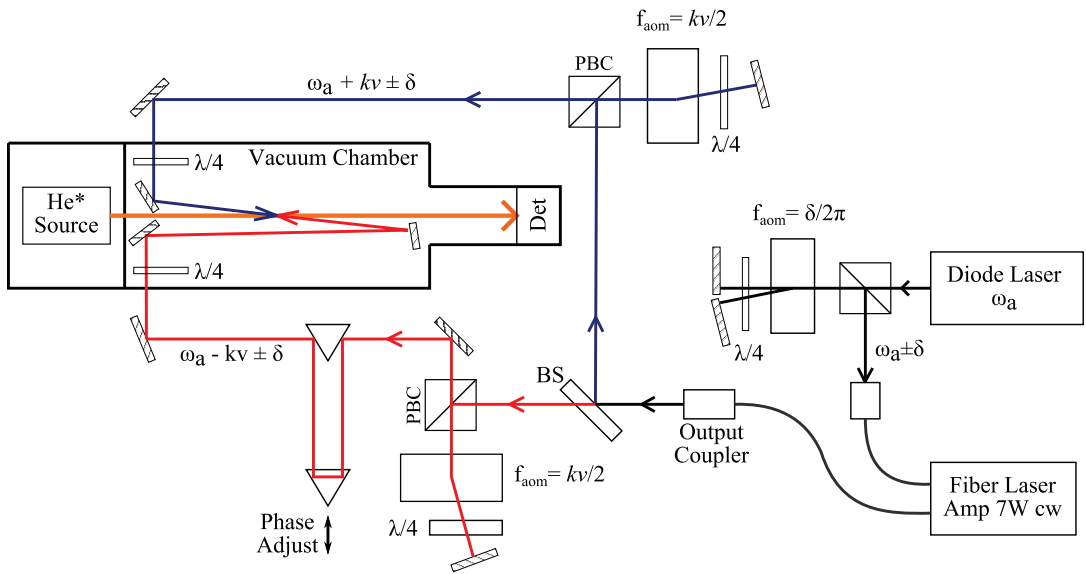}
\caption{(color online).  Experimental configuration for observing the BCF in atomic He with $\sigma^+$- and $\pi$-polarized light.  The $\lambda/4$ retarders inside the vacuum chamber are removed to produce $\pi$ polarization.  Polarizing beamsplitter cubes are denoted by PBC, and a beamsplitter by BS.}
\label{fig:explayout}
\end{figure*}

A diagram of the experimental apparatus is shown in Fig.~\ref{fig:explayout}. The metastable helium source is a reverse-pumped, liquid nitrogen cooled dc discharge source designed by Kawanaka {\itshape et. al.}~\cite{Kawanaka93} as modified by the group of Niehaus~\cite{Mastwijk98}, with additional external modifications to permit installation in an existing vacuum chamber.  The metastable atom flux from the source is about $10^{11}$~He* atoms/sr$\cdot$s with a most probable velocity of approximately 1050~m/s and a metastable fraction of roughly $5\times10^{-5}$~\cite{Metcalf99}. The atom beam passes from the source chamber through a 500~$\mu$m diameter skimmer aperture into a time-of-flight chamber in which it is collimated by a 70~$\mu$m diameter collimating aperture before being mechanically chopped by a tuning fork chopper operating at 160 Hz with a 100~$\mu$m slit width.  The resulting chopped atom beam, with a divergence half-angle of 4.1 mrad, travels through the counterpropagating BCF decelerator lasers for a few cm.  After an additional time-of-flight path, the chopped beam impinges on a stainless steel Faraday cup.  Electrons ejected from the surface by He* atom impacts are collected and detected by a Ceratron continuous dynode electron multiplier, and the He* velocity distribution is then calculated from the time of flight spectra.

\begin{figure}
\includegraphics[width=0.9\columnwidth]{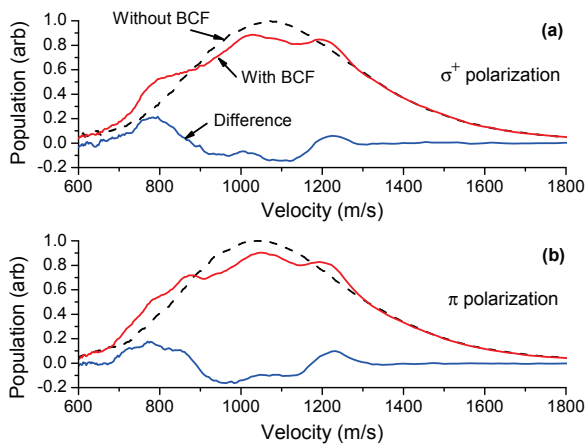}
\caption{(color online).  Comparison of bichromatic slowing of metastable helium using $\sigma^{+}$ polarization (top) and $\pi$ polarization at twice the laser irradiance (bottom).  Effects of the BCF are observed between about 700 and 1050 m/s, but are confined to about 20\% of the atoms within the BCF velocity range, because of limited transverse overlap with the tightly focused lasers.  The additional shifts in the velocity profile at 1050-1300 m/s are not bichromatic in nature --- they arise from the strongly saturated force from one of the four component beams acting alone, producing small shifts for a large number of atoms.}
\label{fig:result}
\end{figure}

The bichromatic decelerator beams are generated from a single diode laser locked to the $2\,^{3}S_{1} - 2\,^{3}P_{2}$ transition at 1083.3~nm using a saturated absorption spectrometer. The laser at frequency $\omega_{12}$ (see Fig. \ref{fig:helevels}) is double-passed through an acousto-optic modulator (AOM) operating at the bichromatic detuning frequency $\delta/2\pi = 300$ MHz producing two superimposed frequencies at $\omega_{12}\pm \delta$ and then coupled into a fiber laser amplifier with a single-mode output of up to 7~W cw. The amplified beatnote train is split into equal parts and Doppler frequency shifts of $\pm kv/2 \pi = \pm$800~MHz are added using additional AOMs to center the bichromatic force at about 866~m/s. An optical delay line is used to set the relative phase between the two bichromatic beat note trains to the optimal $\phi = \pi/2$~\cite{Soding97}. The bichromatic beams enter the chamber linearly polarized, pass through optional quarter-wave retarders to produce $\sigma^{+}$ light when required, and cross the helium atomic beam at an angle of about $1^{\circ}$. The beams are tightly focused to waists with top-hat radii of 0.32~mm to provide the required irradiance without exceeding the capabilities of the amplifier, resulting in an overlap with the atomic beam about 3.1~cm long but with a transverse overlap of only about 20\%.

Initially the bichromatic slowing parameters were optimized using $\sigma^+$ light and time-of-flight profiles were acquired, then the \textit{in vacuo} $1/4$~wave retarders were removed and the experiment was repeated with $\pi$ polarization at twice the irradiance, under otherwise identical conditions.  Finally, the retarders were replaced and the $\sigma^+$ experiment was run a second time to verify reproducibility of the results.

\subsection{Experimental results\label{res}}

The resulting velocity distributions of the metastable helium atoms are shown in Fig.~\ref{fig:result}. Because only 20\% of the beam area overlapped the central full-power region of the Gaussian laser profile, the changes due to the BCF are limited in amplitude.  In both experimental configurations, a corresponding fraction of the atoms are slowed by an average of $\Delta v \cong 180$~m/s.  This corresponds to an average slowing of $100\gamma \approx \delta / 2k$, about 1/2 the full velocity range of the bichromatic force~\cite{Cashen01}, as expected.
The total atom number in both cases is preserved to better than 1\%. Comparison of the velocity distributions in parts (a) and (b) of Fig.~\ref{fig:result} shows that the effects of the BCF are quite similar for $\sigma^+$ and $\pi$ polarization, verifying our expectations.  Although we cannot accurately measure the force reduction for the $\pi$-polarized case, we can say with assurance that it is by no more than a factor of $3-4$.  This demonstration confirms that the bichromatic force can successfully be used in slowing atoms with more than one cycling transition.

\begin{table}
\caption{\label{tbl:He*_shifted}Measured BCF depletion percentages, as a function of the frequency shift of $\omega_0$ from exact atomic resonance.  Shifts are measured as a fraction of the bichromatic detuning $\delta$.  Because the interaction length for this measurement was 3-4 times the BCF slowing distance, little effect is expected until the force is reduced by a corresponding factor.}
\begin{ruledtabular}
\begin{tabular}{cc}
Fractional Shift &	Depletion (\%)\\
\hline
0 & 35\\
0.1 & 30\\
0.2 & 27\\
0.3 & 29\\
0.4 & 16\\
\end{tabular}
\end{ruledtabular}
\end{table}

In a separate variation of the experiment, we have also specifically tested the effects of frequency shifts in the pure two-level configuration, using an arrangement slightly different from Fig. \ref{fig:result}.  For this measurement a small bichromatic detuning $\delta = \pm 100$ MHz was used so that the fiber amplifier could be omitted.  The bichromatic slowing parameters were experimentally optimized with $\sigma^+$ light, and then the laser center frequency $\omega_0$ was offset by using an AOM in each of the two-color beams.  We characterized the efficiency of BCF slowing by measuring the depletion dip in the velocity profile, at the velocity that corresponding to the minimum of the dip for zero shift.  Table \ref{tbl:He*_shifted} shows the results, which show only minor effects until the shift reaches 0.4 $\delta$.  This is consistent with the calculated force reduction factors $\eta$ in Table \ref{tbl:EtaVals}, because the interaction length was about 3-4 times the BCF slowing distance, meaning that the velocity profiles should not be much affected until the force is reduced by a factor of 3-4.  Table \ref{tbl:EtaVals} indicates that this should occur for shifts exceeding about 0.3 $\delta$, where a sudden decline in the depletion dip is indeed observed.  The most important aspect of this test, though, is not the onset of the decline so much as the demonstration that for shifts below 0.3 $\delta$ the BCF works well.  This explicitly verifyies our expectation that the BCF will be robust against small molecular frequency splittings.

\section{Summary\label{sec:Summary}}
It appears practical to use the BCF to obtain order-of-magnitude enhancements over the radiative force on diatomic molecules.  The necessary conditions are that ample laser power is available, that the small-scale spectroscopic structure due to fine and hyperfine structure is not excessively complex, and that the rotational branch selected for BCF excitation is well-isolated from other branches involving the same upper and lower levels.   Specific calculations for the $A-X$ transition in CaF show considerable promise for BCF deceleration and cooling on the $Q_{11}$ branch of the (0-0) band, including the ability to decelerate an initially cryogenic beam to a stop if a vibrational repumping laser is used.  Operation with faster initial velocities may be possible by incorporating refinements such as properly phased multipass beams and coherent repumping.  We demonstrate with an atomic helium experiment that the BCF is not greatly impacted by the presence of multiple, simultaneously-cycling transitions, by using $\pi$-polarized light to simultaneously drive three magnetic sublevels of the $2\,^3P_2-2\,^3S_1$ transition.  In this example a substantial BCF is obtained even in the presence of strong off-resonant coupling to the nearby $2\,^3P_1$ state, which produces large level-dependent frequency shifts.

We also remark that the conditions for significant transverse deflection and cooling are much less stringent than for longitudinal deceleration, and the BCF has considerable promise as a method for producing state-selected molecular beams by deflection using long-pulsed lasers, even for transitions that are not close to meeting the near-cycling condition.

\begin{acknowledgments}
We thank David DeMille, John Barry, and Edward Shuman for several helpful discussions.  Financial support was provided by the University of Connecticut Research Foundation, the National Science Foundation, and the Air Force Office of Scientific Research (MURI FA9550-09-1-0588).
\end{acknowledgments}


\end{document}